# Determination of the phase coherence length of PdCoO$_2$ nanostructures

T. Harada[1*], P. Bredol[2], H. Inoue[1,3], S. Ito[1], J. Mannhart[2], and A. Tsukazaki[1,4]

[1] *Institute for Materials Research, Tohoku University, Sendai 980-8577, Japan*
[2] *Max Planck Institute for Solid State Research, Heisenbergstr. 1, 70569 Stuttgart, Germany*
[3] *Frontier Research Institute for Interdisciplinary Sciences, Tohoku University, Sendai 980-8578, Japan*
[4] *Center for Spintronics Research Network (CSRN), Tohoku University, Sendai 980-8577, Japan*
*\*Correspondence to: t.harada@imr.tohoku.ac.jp*

The two-dimensional layered compound PdCoO$_2$ is one of the best oxide conductors, providing an intriguing research arena opened by the long mean free path and the very high mobility of ~ 51000 cm$^2$/Vs. These properties turn PdCoO$_2$ into a candidate material for nanoscale quantum devices. By exploring universal conductance fluctuations originating at nanoscale PdCoO$_2$ Hall-bar devices, we determined the phase coherence length of electron transport in *c*-axis oriented PdCoO$_2$ thin films to equal ~ 100 nm. The weak temperature dependence of the measured phase coherence length suggests that defect scattering at twin boundaries in the PdCoO$_2$ thin film governs phase breaking. These results suggest that phase coherent devices can be achieved by realizing the devices smaller than the size of twin domains, via refined microfabrication and suppression of twin boundaries.

Quantum phase coherence in mesoscopic conductors has been intensively studied to explore fundamental questions of quantum mechanics as well as to pursuit novel device architectures [1,2]. The phase coherence length $l_\phi$ of the charge carriers is the fundamental key parameter governing the quantum interference phenomena in such mesoscopic devices. This coherence length is a measure of the distance over which an electron propagates while maintaining its phase information. The dominant origin of the phase breaking of conducting charges is inelastic scattering provided by electron-electron, electron-phonon, and electron-defect scattering [3]. Information on $l_\phi$ is obtainable from studies of universal conductance fluctuations (UCF). In a conductor with a size smaller or comparable to $l_\phi$, we can expect the electron interference resulting from travel on different trajectories ($\gamma_n$ and $\gamma_m$ in Fig. 1(a)). Being dependent on interference patterns of electron wavefunctions, the total conductance of the channel fluctuates due to rearrangement of scattering sources as well as phase shifts induced by magnetic fields [2,4,5]. By analyzing the UCF in magnetoconductance, phase coherence length can be precisely evaluated.

The highly conductive layered metal PdCoO$_2$ has a unique anisotropic crystal structure with alternating Pd$^+$ and [CoO$_2$]$^-$ layers (Fig. 1(a), left) [6,7]. Whereas the Pd$^+$ layers mediate the electron conduction [8,9], the [CoO$_2$]$^-$ layers are of insulating nature, forming two-dimensional (2D) electronic systems. In fact, a cylindrical Fermi surface with nearly hexagonal cross-section has been observed by angle-resolved photoemission spectroscopy (ARPES) [10] and the de Haas-van Alphen effect [9]. The closed Fermi surface geometry minimizes the effect of electron-phonon and Umklapp scattering processes [9], as is the case in alkaline metals [11]. The high conductivity with the long electron mean free path (~ 20 μm) [9,12-14] reported for a bulk single crystal makes PdCoO$_2$ a promising platform for studying quantum transport [15]. As explored in semiconductor heterostructures in the last decades [2], the quantum interference effects have been intensively studied with mesoscopic devices fabricated by a well-regulated growth technique and high-resolution lithography techniques. As for PdCoO$_2$, *c*-axis oriented thin films have been achieved by pulsed-laser deposition (PLD) [16,17], molecular beam epitaxy [18,19], and solid-phase reactions of precursors [20,21]. Establishing a route to pattern PdCoO$_2$ thin films to submicron scales is essential for realizing quantum devices utilizing PdCoO$_2$ thin films and heterostructures. In this study, we report on the determination of the phase coherence length of conducting electrons in mesoscopic Hall-bar devices of PdCoO$_2$ thin films by analyzing UCF. Based on the autocorrelation analysis of the UCF, we suggest that twin boundaries in the films are one of the dominant scattering sources that cause phase breaking in the PdCoO$_2$ nanostructures.

A *c*-axis oriented PdCoO$_2$ thin film with the thickness $d = 6.8$ nm was grown by PLD on an Al$_2$O$_3$ (0001) substrate (Fig. S1). The PdCoO$_2$ thin film was patterned into mesoscopic Hall-bar devices as shown in Fig. 1(b) using electron-beam lithography and Ar-ion milling [22]. A negative resist composed of hydrogen silsesquioxane (HSQ) was used as a mask for Ar-ion milling. Figure 1(c) shows the scanning electron microscope (SEM) image of the HSQ mask patterned on the PdCoO$_2$ thin film before Ar-ion milling. According to the SEM image, the width $W$ and the length $L$ of the Hall-bar device were estimated to be $W = 93$ nm and $L = 410$ nm, respectively (Fig. 1(b) and (c)). The triangular shapes in Fig. 1(c) are attributed to the surface morphology of the PdCoO$_2$ film [16]. The longitudinal ($V_{xx}$) and the transverse voltage ($V_{yx}$) were measured by a lock-in technique using alternating excitation current ($I_{ac}$).

The temperature ($T$) dependence of resistivity ($\rho_{xx} = V_{xx}Wd/I_{ac}L$) showed metallic behavior down to $T \sim 50$ K (Fig. 1(d)), below which the $\rho_{xx}$ was almost constant. The Hall resistivity ($\rho_{yx} = V_{yx}d/I_{ac}$) displayed a linear magnetic field ($\mu_0 H$) dependence with a negative slope, consistent with the electrical conduction being dominated by electron-type charge carriers [10] (Fig. 1(e)). The carrier density ($n$) evaluated by the Hall effect measurement was almost constant in the measured temperature range below 42 K, being around $n = 1.5 \times 10^{22}$ cm$^{-3}$ (Fig. 1(f)). This value is comparable with the bulk value of $2.45 \times 10^{22}$ cm$^{-3}$ (Ref. [6]). The transport mean free path of the electrons ($l_{mf}$) is readily estimated using the Drude model $l_{mf} = v_F \tau = m^* v_F / ne^2 \rho_{xx}$, where $v_F$ is Fermi velocity, $\tau$ is scattering time, $m^*$ is effective mass of electrons, and $e$ is elementary charge [23]. With applying $m^* = 1.49 m_0$ and $v_F = 7.5 \times 10^5$ ms$^{-1}$ as reported for bulk single crystals [6], where $m_0$ is the mass of rest electrons, mean free path $l_{mf}$ is estimated to be approximately 10 nm below 42 K as plotted in Fig. 1(f). From the relation $l_{mf} < W$ and $L$, we conclude that the electron transport in the device is in the diffusive regime as shown in the right schematics of Fig. 1(a).

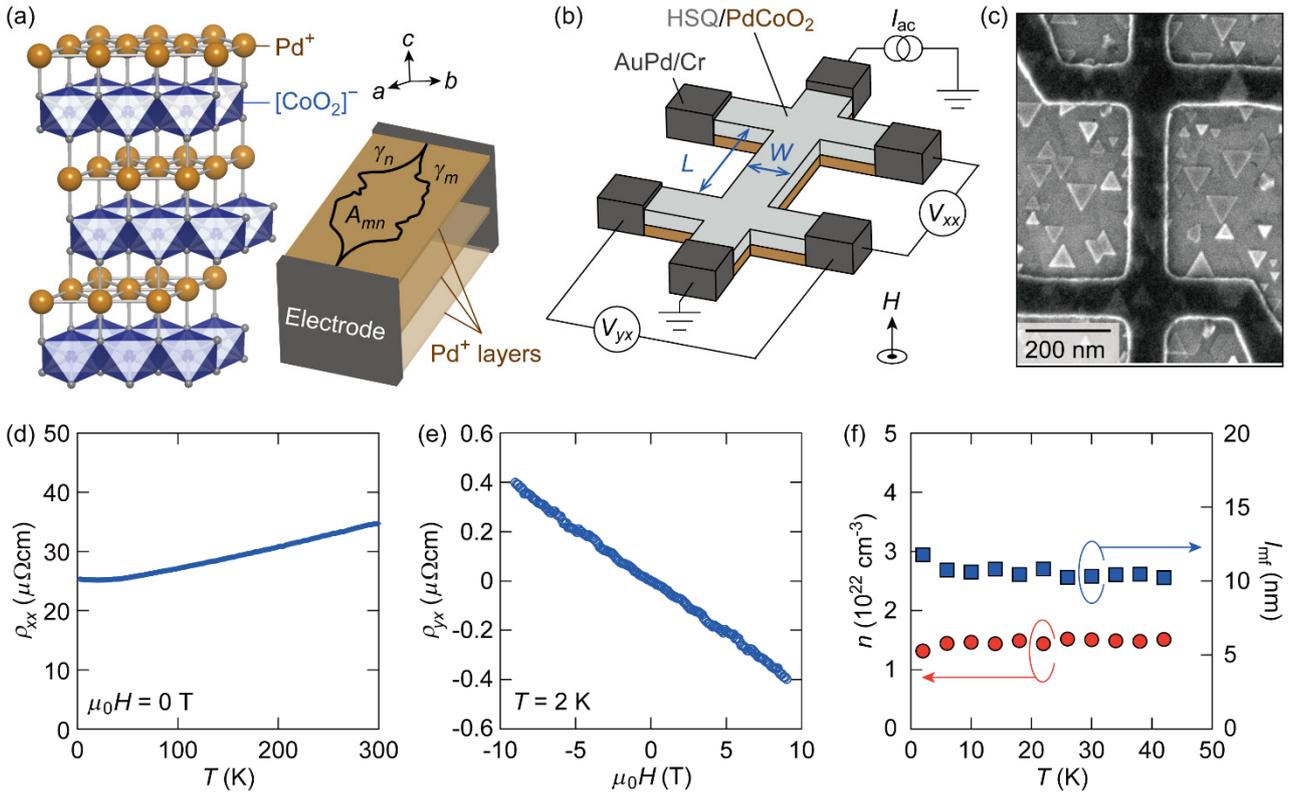

FIG. 1. (Color online) (a) Left: the crystal structure of PdCoO$_2$. Right: a schematic drawing of PdCoO$_2$ channel with Pd$^+$ conductive sheets connected to the electrodes. Two trajectories of electrons, $\gamma_n$ and $\gamma_m$, are shown as black curves. The area surrounded by $\gamma_n$ and $\gamma_m$ is noted as $A_{mn}$. A magnetic flux penetrating $A_{mn}$ alters the phase difference of the trajectories $\gamma_n$ and $\gamma_m$. (b) A schematic of a Hall-bar device fabricated on c-Al$_2$O$_3$ substrates. The longitudinal ($V_{xx}$) and transverse voltage ($V_{yx}$) were measured using the alternating excitation current ($I_{ac}$) under a magnetic field $H$ applied perpendicular to the PdCoO$_2$ top surface. $L$ and $W$ stand for the separation between the voltage terminals and the width of the channel, respectively. (c) A SEM image of the HSQ resist (dark region) patterned on PdCoO$_2$/c-Al$_2$O$_3$. The triangular patterns of surface morphology were also visible. (d) Resistivity ($\rho_{xx}$) versus $T$ properties of the Hall-bar device under $\mu_0 H = 0$ T. (e) Hall resistivity ($\rho_{yx}$) versus $\mu_0 H$ data measured at $T = 2$ K. (f) Temperature dependence of the carrier density ($n$) (red circles) and the mean free path ($l_{mf}$) (blue squares) estimated by the Drude model: $l_{mf} = v_F \tau = m^* v_F / n e^2 \rho_{xx}$, where $v_F$ is Fermi velocity, $\tau$ is scattering time, $m^*$ is the effective mass of electrons, and $e$ is the elementary charge. We used $m^* = 1.49 m_0$ and $v_F = 7.5 \times 10^5$ ms$^{-1}$ taken from measurements of bulk samples [6].

Under perpendicular magnetic field ($\mu_0 H$) at $T = 2$ K, the $\rho_{xx}(\mu_0 H)$ dependence shows characteristic fluctuations (Fig. 2(a), red) superposed on the negative magnetoresistance. The amplitude of the fluctuations decreases above $T = 10$ K (green) and vanishes at $T \sim 26$ K (blue). Such fluctuations of $\rho_{xx}$ have been observed in mesoscopic structures of metals [24,25], semiconductors [26-29], graphene [30], and topological insulators [31-33] as a result of quantum interference effect. The amplitude of the fluctuation in a micrometer-sized device of the PdCoO$_2$ film is much smaller than that in a mesoscopic device (Fig. S2) [22], indicating that the fluctuation originates from mesoscopic phenomena. To analyze the fluctuations quantitatively, the channel conductance $G(T, \mu_0 H) = I_{ac} / V_{xx} = Wd / L\rho_{xx}$ has to be considered. The background $G_0(T, \mu_0 H) = Wd / L\rho_{xx0}$ is subtracted to extract the conductance fluctuation $\delta G = G(T, \mu_0 H) - G_0(T, \mu_0 H)$. Here $\rho_{xx0}$ is the $\mu_0 H$-dependent smoothed resistivity curve, plotted as black dashed line for $T = 2$ K in Fig. 2(a). The $\delta G$ versus $\mu_0 H$ characteristics, shown in Fig. 2(b), exhibit broad fluctuations with dips around $\mu_0 H = 4$ T and 7 T reproducibly from $T = 2$–26 K. As shown for $T = 2$ K, the dip features are consistently observed in both sweep directions of magnetic field (red and black lines in Fig. 2(b)), thereby differing from extrinsic random noise.

We analyze the aperiodic fluctuations shown in Fig. 2(b) using the standard UCF model [5,34]. Here, we introduce the autocorrelation function $F(T, \mu_0 \Delta H)$ for $\delta G$ as

$$F(T, \mu_0 \Delta H) = \langle \delta G(T, \mu_0 H) \delta G(T, \mu_0 H + \mu_0 \Delta H) \rangle_{\mu_0 H},$$

where $\langle ... \rangle_{\mu_0 H}$ stands for averaging over $\mu_0 H$. As plotted in Fig. 3, the $F(T, \mu_0 \Delta H)$ dependence changes systematically as the temperature is increased from 2 to 30 K. The correlation field ($\mu_0 H_c$) is evaluated by using the relation $F(T, \mu_0 H_c) = F(T,0)/2$ (ref. [34]), as plotted in the inset of Fig. 3. $H_c$ corresponds to a magnetic flux in the phase coherent region of the order of a flux quantum $\phi_0 = h/e$. Thus, the $\mu_0 H_c$ is related to the phase coherence length $l_\phi$ as $\mu_0 H_c = \beta_1 \phi_0/W l_\phi$ for 1D systems ($l_\phi \gg W$) and $\mu_0 H_c = \beta_2 \phi_0/l_\phi^2$ for 2D systems ($l_\phi \ll W$), where $\beta_1$ and $\beta_2$ are geometry-dependent constants of order unity [34,35]. Applying these formulas, the phase coherence length is estimated to be approximately 100 nm at 2–20 K as plotted in Fig. 4(a). We use the symbols $l_\phi^{1D}$ and $l_\phi^{2D}$ to distinguish the phase coherence length estimated by the 1D and 2D models, respectively. As both values of $l_\phi^{1D}/\beta_1$ (blue circles) and $l_\phi^{2D}/\beta_2$ (green squares) are close to the device width $W$ as shown in Fig. 4(a), the dimension of the system is likely in the crossover regime between 1D and 2D. Although according to the standard model [36], $l_\phi$ is expected to decay with $T^{-1/3}$ in 1D and $T^{-1/2}$ in 2D systems, both $l_\phi^{1D}/\beta_1$ and $l_\phi^{2D}/\beta_2$ do not depend much on $T$. The linear fitting of the logarithmic plots gives the slope of $-0.18$ for $l_\phi^{1D}/\beta_1$ (blue line) and $-0.09$ for $l_\phi^{2D}/\beta_2$ (green line), the absolute values of which are much smaller than the values predicted by theory ($-0.33$ and $-0.5$, respectively). On the other hand, the root mean square of the conductance fluctuation, calculated as

$$rms(\delta G) = \left\{\left\langle \delta G^2 \right\rangle_{\mu_0 H}\right\}^{1/2} = \left\{F(T,0)\right\}^{1/2},$$

decreases with increasing temperature (Fig. 4(b)). The $rms(\delta G)$ represents a $T^{-1/2}$ dependence (red line) from the largest value of 0.07 $e^2/h$ at $T = 2$ K. Such a temperature dependence of the $rms(\delta G)$ often comes from the temperature dependence of $l_\phi$ [34]. As the $l_\phi^{1D}/\beta_1$ and $l_\phi^{2D}/\beta_2$ are hardly dependent on temperature, however, it is not the phase coherence length that causes the temperature dependence of the $rms(\delta G)$ of the PdCoO$_2$ Hall-bar device.

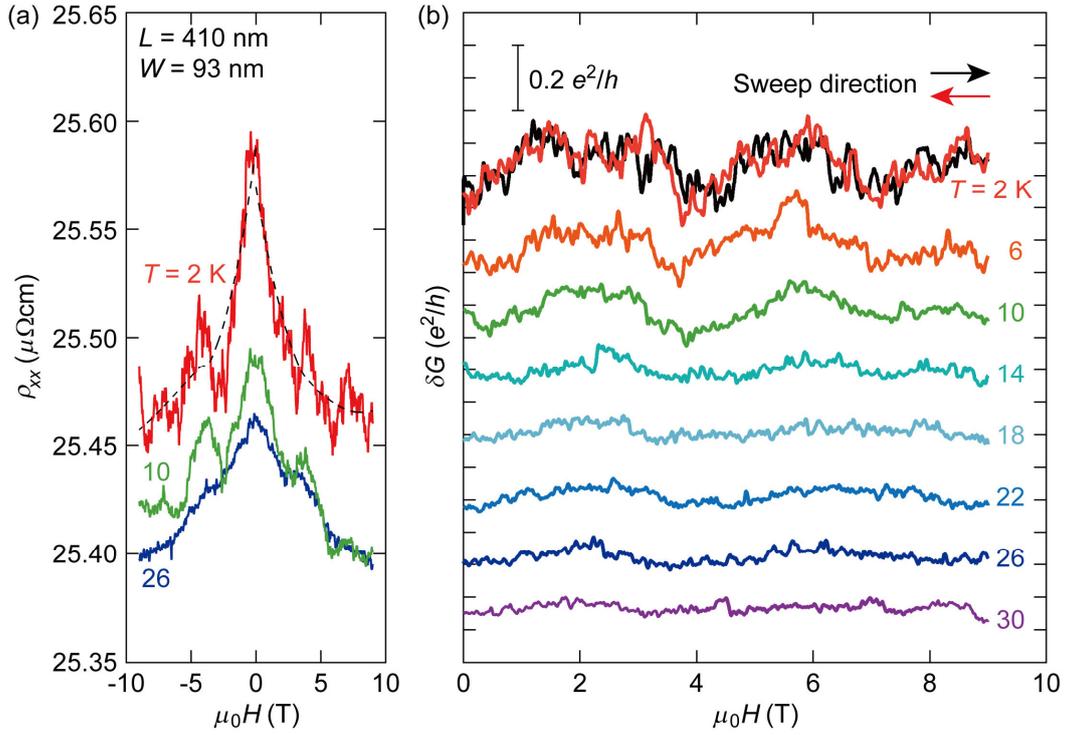

**FIG. 2.** (Color online) (a) The measured $\mu_0 H$ dependence of $\rho_{xx}$ at $T = 2$, 10, and 26 K. The amplitude of $I_{ac}$ was set to be $I_{ac} = 500$ nA. The smoothed background ($\rho_{xx0}$) for $T = 2$ K is shown as the black dashed line. (b) Conductance fluctuation $\delta G$ obtained by subtracting the smoothed background from the $I_{ac}/V_{xx}(\mu_0 H)$ curves for $T = 2$–30 K. The $\delta G$ curves are offset along the vertical axis for clarity. The $\delta G$ of 0.2 $e^2/h$ is shown as a scale for the vertical axis. For $T = 2$ K, two curves measured by sweeping $\mu_0 H$ from 0 to 9 T (black) and from 9 T to 0 T (red) are shown. The curves for the other temperatures were measured by sweeping $\mu_0 H$ from 9 T to 0 T.

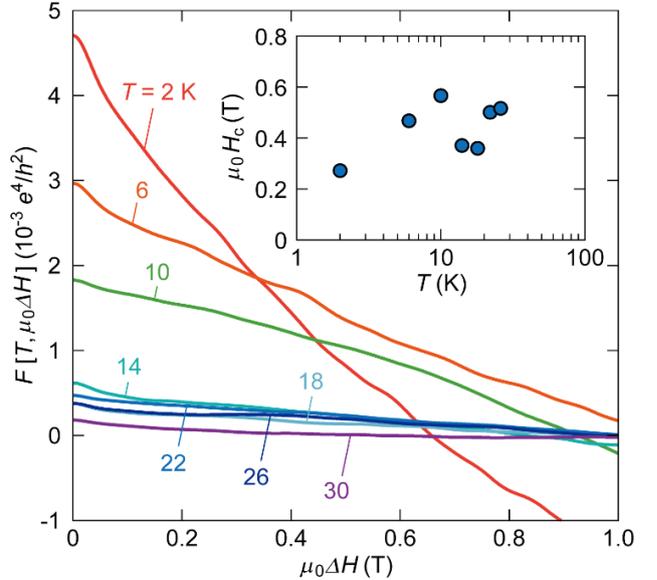

**FIG. 3.** (Color online) The autocorrelation function $F(T, \mu_0 \Delta H) = \langle \delta G(T, \mu_0 H) \delta G(T, \mu_0 H + \mu_0 \Delta H) \rangle_{\mu_0 H}$ plotted as a function of $\mu_0 \Delta H$, obtained from the conductance fluctuations measured at the temperatures noted. The inset displays the temperature dependence of the correlation field $\mu_0 H_c$.

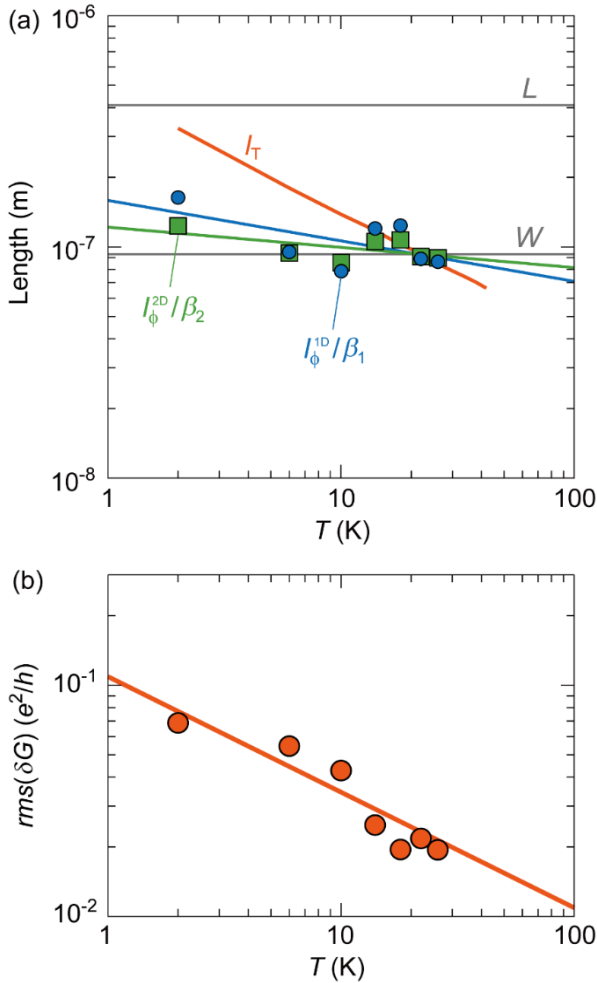

FIG. 4. (Color online) (a) Temperature dependence of $l_\phi^{1D}/\beta_1$ (blue circles), $l_\phi^{2D}/\beta_2$ (green squares), and the thermal length ($l_T$) (red line). The thermal length $l_T$ is estimated using $l_T = (\hbar D/k_B T)^{1/2}$ and $D = v_F^2 \tau/2 = v_F^2 m^* \rho_{xx}^{-1}/2ne^2$, where $D$ is the electronic diffusion constant [23,34]. For the calculation of $l_T$ and $D$, we used the $\rho_{xx}$ and $n$ plotted in Fig. 1(d) and Fig. 1(f), respectively. The gray lines correspond to the width $W$ and the length $L$ of the Hall-bar device. The blue and green lines are the linear fitting to $l_\phi^{1D}/\beta_1$ and $l_\phi^{2D}/\beta_2$, respectively. (b) The root mean square of the conductance fluctuation $rms(\delta G) = F[0]^{1/2}$. The red line shows the $T^{-1/2}$ dependence.

We therefore interpret the $T^{-1/2}$ dependence of $rms(\delta G)$ as thermal averaging effect that is characterized by the thermal length $l_T = (\hbar D/k_B T)^{1/2}$ (Fig. 4(a)), which is comparable to $l_\phi^{1D}/\beta_1$ and $l_\phi^{2D}/\beta_2$.

It is noteworthy that the evaluated phase coherence length barely changes with temperature. This fact strongly suggests that the phase breaking is caused by a temperature-independent effect, which is likely related to the defects present in the device [37]. The boundaries of the twin domains of the PdCoO$_2$ thin films are candidates of such defects. The orientations of triangular step-and-terrace structures (Fig. 5(a) and Fig. S3) can depend on the in-plane orientation of 180°-rotated crystal twins [16] and/or the surface energy of the growing domains. To therefore analyze the twin boundaries in the PdCoO$_2$ samples, we applied high-angle annular dark-field scanning transmission electron microscope (HAADF-STEM) (Fig. 5(b)). The whitish dots in Fig. 5(b) correspond to the Pd atoms with the large atomic number in the Z-contrast image. The layered structure in the image reflects the c-axis orientation of the film. Regarding the orientations in the ab-plane, we can classify twin domains by the resulting difference of the lattice arrangements along the Pd-Co-Pd columns highlighted with black lines. There are twin boundaries (red arrow) and stacking faults (green arrow) caused by the twinning. Considering the nearly 2D conduction properties of PdCoO$_2$ [9], the twin boundaries as indicated by the red arrow mainly influence on the in-plane conduction. Although it is difficult to detect multiple twin boundaries in one micrograph, we can determine the minimum distance of the twin boundaries in our experiment to equal at least 40 nm. The density of the twin boundaries is therefore in the range to account for the $l_\phi^{1D}/\beta_1$ and $l_\phi^{2D}/\beta_2$ of about 100 nm (Fig. 4(a)).

The periodicity of the Pd lattice at the twin boundaries is schematically depicted in Fig. 5(c). As shown by the upper triangle, the expected Pd sites (black dashed line) of the upper domains are located at the interstitial sites of the other twin. Thus, the twin boundaries can be regarded as planes consisting of interstitial Pd atoms and/or Pd vacancies. Such planes induce electron scattering. Indeed, Frenkel pairs, i.e. combinations of an interstitial Pd and a Pd vacancy, were reported to significantly increase the resistivity of PdCoO$_2$ single crystals [38]. The long coherence length within the domains can be bounded by the inelastic scattering at the twin boundaries, where a slight mismatch of the Fermi surface between the twins may play a role. The disappearance of the conductance fluctuation above 26 K is explained by $l_T$ becoming sufficiently shorter than the size of twin domains (> 40 nm) (Fig. 4(a)). According to these considerations, the twin boundaries are likely the dominant cause of phase breaking in the PdCoO$_2$ thin film.

In summary, we have measured the electrical transport properties of a submicron-scale Hall-bar device of a PdCoO$_2$ thin film. Universal conductance fluctuations are found in magnetoresistance at temperatures below 26 K. By applying an autocorrelation analysis, the phase coherence length of the electrons in the PdCoO$_2$ Hall-bar device is found to equal ~ 100 nm at 2 K. The phase breaking length is proposed to be limited by the existence of crystal twin boundaries that cause phase-breaking scattering of the conduction electrons. This demonstration of quantum coherence in a PdCoO$_2$ nanostructure is a first step to study the interplay of quantum transport and the exotic properties caused by the high conductivity [13,14] and the polar surface [39-41] of PdCoO$_2$ in thin-film mesoscopic devices. According to our results, the suppression of twin boundaries is essential for further extending the phase coherence length. Such suppression may be possible by use of delafossite-type substrates [42] with optimized miscut angles [17] to lift the degeneracy of the formation energy of twins.

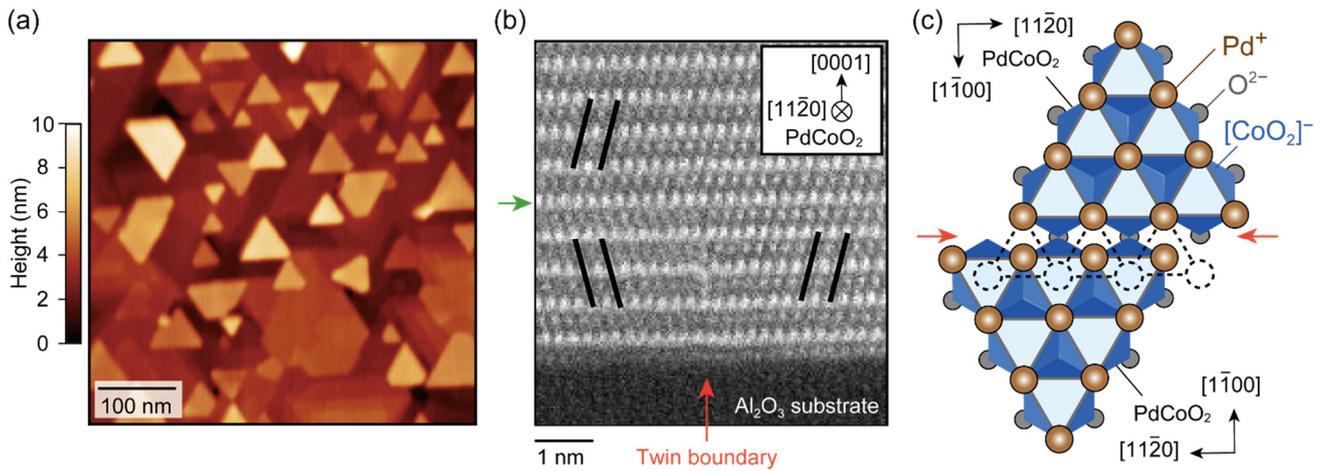

**FIG. 5.** (Color online) (a) The AFM image of a PdCoO$_2$ thin film ($d$ = 6.8 nm) grown on an Al$_2$O$_3$ (0001) substrate. (b) The HAADF-STEM image of a PdCoO$_2$ thin film ($d \sim$ 10 nm). The twin domains can be classified by the arrangement of the Pd-Co-Pd columns (highlighted with the black lines). A twin boundary and a stacking fault are marked by red and green arrows, respectively. (c) The crystal structure at a twin boundary (red arrows). The black dashed line indicates the positions of Pd atoms of the extended crystal lattice of the upper twin.

### Acknowledgements

We thank K. Fujiwara for valuable comments on the manuscript. This work is a cooperative program (Proposal No. 18G0407 and No. 19K23415) of the CRDAM-IMR, Tohoku University. This work is partly supported by a Grant-in-Aid for Early-Career Scientists (No. 18K14121) from the Japan Society for the Promotion of Science (JSPS) and JST CREST (JPMJCR18T2), Mayekawa Houonkai Foundation, Tanaka Kikinzoku Memorial Foundation, and Grant for Basic Science Research Projects from The Sumitomo Foundation.### References